# Which gravitomagnetic precession rate will be measured by Gravity Probe B?

Jacob Biemond*

*Vrije Universiteit, Amsterdam, Section: Nuclear magnetic resonance, 1971-1975*

## Abstract

General relativity predicts a "hyperfine" precession rate for a gyroscope moving in the gravitomagnetic field of a rotating massive body. The recently launched Gravity Probe B (GP-B) will test the predicted precession rate of 40.9 milliarc-seconds per year for a set of four gyroscopes in a Polar Earth Orbit (PEO).

It may be possible, however, that the gravitomagnetic field from a rotating mass behaves in the same way as the magnetic field generated by a moving charge. In that case the predicted precession rate of a gyroscope will be zero, since the gyroscopes of GP-B have been shielded against external magnetic fields.

Another possible manifestation of the equivalence of gravitomagnetic and magnetic field may already have been found. It is the so-called Wilson-Blackett law, approximately describing the magnetic field of many rotating celestial bodies.

In this work a review of the gravitomagnetic approach is given starting from the Einstein equations. Four gravitomagnetic equations, analogous to the Maxwell equations, are deduced. The Wilson-Blackett relation follows from these equations, if the gravitomagnetic field is identified as a common magnetic field.

In addition, the precession rate for a gyroscope in terms of the gravitomagnetic field has been derived, starting from the principle of general covariance. The gravitomagnetic field may again be identified as a common magnetic field, or can be evaluated in the standard way. The future observations from GP-B may discriminate between the alternative choices.

## 1. Introduction

One of the aims of the Gravity Probe B [1–3] is to measure the gravitomagnetic precession rate of a gyroscope in a polar orbit about the Earth [4, 5]. In the framework of fixed stars the precession rate $\mathbf{\Omega}$ deduced by Schiff [5] is given by

$$\mathbf{\Omega} = c^{-2} G \left( \frac{3 \, \mathbf{S} \cdot \mathbf{R}}{R^5} \mathbf{R} - \frac{\mathbf{S}}{R^3} \right), \qquad (1)$$

where $\mathbf{S} = I \mathbf{\omega} = 2/5 \, f \, m \, r_0^2 \, \mathbf{\omega}$ is the angular momentum for a large sphere of radius $r_0$, $I$ is its moment of inertia, $\mathbf{\omega}$ its angular velocity, $m$ its mass and $f$ is a dimensionless factor depending on the homogeneity of the mass density in the sphere (for a homogeneous mass distribution $f = 1$). $\mathbf{R}$ is the position vector from the centre of, e.g., the Earth to a gyroscope. For a gyroscope moving in a polar orbit 649 km above the Earth the integrated value of $\Omega$ is equal to 40.9 milliarc-seconds per year (mas/yr). The Gravity Probe B will measure this effect to a precision of 1% or better [1, 3].

As has been discussed by Weinberg [6, chapters 5, 9], equation (1) can be derived from the equations of spin motion dictated by the principle of general covariance. Starting from the same principle but applying the gravitomagnetic approach, the precession rate $\mathbf{\Omega}$ can also be written in terms of the so-called "magnetic-type" gravitational field $\mathbf{B}$ [7–9]

$$\mathbf{\Omega} = -2 \, \beta^{-1} \, c^{-1} \, G^{1/2} \, \mathbf{B}, \qquad (2)$$

where $\beta$ is a dimensionless constant of order unity. Note that the field $\mathbf{B}$ in (2) has the dimension of a magnetic induction field, but alternative dimensions for the "magnetic-type" gravitational field are also mathematically possible.

---

*Postal address: Sansovinostraat 28, 5624 JX Eindhoven, The Netherlands.
Website: http://www.gewis.nl/~pieterb/gravi/, e-mail: gravi@gewis.nl

In our version of the gravitomagnetic theory [8, 9] it is assumed that the gravitomagnetic field **B**(gm) and an electromagnetic field **B**(em) due to moving charge are equivalent. Thus, a gyroscope surrounded by a total field **B**(tot) = **B**(gm) + **B**(em) cannot distinguish between these fields of different origin and **B** = **B**(tot) has to be substituted into (2). However, the gyroscopes in the experimental set-up of GP-B have carefully been shielded against all external magnetic fields [1–3]. Then, the field **B**(gm) as well as a possible contribution **B**(em) both due to the Earth may be filtered out and a precession rate $\Omega \cong 0$ may be found. Usually, it is assumed, however, that the field **B**(gm) has properties totally different from the magnetic induction field **B**(em) generated by moving charge. In that case the standard precession rate (1) may be observed.

In this work the field **B**(gm) for a rotating mass of electrically neutral matter, will separately be deduced, as previously has been done [7–9]. The following dipolar magnetic field **B** = **B**(gm) at distance $R > r_0$ is obtained

$$\mathbf{B} = \frac{3\,\mathbf{M}\cdot\mathbf{R}}{R^5}\mathbf{R} - \frac{\mathbf{M}}{R^3}, \tag{3}$$

where **M** = **M**(gm) is the gravitomagnetic dipole moment given by

$$\mathbf{M} = -\tfrac{1}{2}\,\beta\,c^{-1}\,G^{\tfrac{1}{2}}\,\mathbf{S}. \tag{4}$$

Note that **M** and **S** possess opposite directions for the choice $\beta = +1$. Substitution of (4) into (3), followed by insertion of the result into (2) yields the standard result for $\Omega$ of (1). It is noticed that the result (1) does not depend on the constant $\beta$.

If the gravitomagnetic field **B** = **B**(gm) of (3) is identified as a common magnetic induction field, expression (4) represents the so-called Wilson-Blackett law. This relation appears to be approximately valid for *many*, *strongly different*, celestial bodies and some rotating metallic cylinders in the laboratory as well (see for a review [9] and references therein; for pulsars, see [10]). It is noticed that the values of **M**(obs) and **S** vary over the large interval of about *sixty decades!* The correct order of magnitude of the observed magnetic dipole moment **M**(obs) compared with **M**(gm) from (4) is the main reason to propose that **B**(gm) from (3) is equivalent to the magnetic induction field **B**(em) generated by moving charge.

In section 2 we first review the deduction of the gravitomagnetic field **B** of (3), whereas in section 3 the precession rate $\Omega$ of (2) is derived and discussed. In section 4 the main results are summarized and conclusions are drawn.

## 2. The gravitomagnetic field

The Einstein equations (without cosmological constant) are the starting point of our gravitomagnetic approach. In this work the formalism given by Landau and Lifshitz [11, chapters 10–13] has been followed. The full Einstein equations are

$$R_{ij} - \tfrac{1}{2}\,g_{ij}\,R = 8\pi c^{-4}\,G\,T_{ij}. \tag{5}$$

In these equations the tensor components $R_{ij}$ and $T_{ij}$ and the invariant $R$ have their usual meaning. In this work it will be assumed that a small mass density $\rho$ moves with a low velocity **v** ($v \ll c$) through a weak gravitational field.

The components of the metric tensor $g_{ij}$ are assumed to be symmetric and can be written as

$$g_{ij} = \eta_{ij} + h_{ij}, \tag{6}$$

where $\eta_{ij} = (1, -1, -1, -1)$. In the applied weak field approximation the metric components $g_{ij}$ are close to the Minkowski metric components $\eta_{ij}$, so $|h_{ij}| < 1$. The following metric written in Cartesian coordinates will be used in this work (Gaussian units are chosen throughout this paper)

$$c^2 d\tau^2 = g_{00} c^2 dt^2 + 2 g_{0\alpha} c\, dt\, dx^\alpha + g_{\alpha\beta} dx^\alpha dx^\beta, \tag{7}$$

where $\tau$ is the "proper time" and $\alpha$ and $\beta$ are summed from 1–3 (= $x, y, z$).

Equations (5) will now be evaluated by introducing a new, mixed tensor $\psi^j_i$ defined by

$$\psi^j_i \equiv h^j_i - \tfrac{1}{2} \delta^j_i h, \tag{8}$$

where $h^j_i \equiv \eta^{jk} h_{ki} = \eta^{jj} h_{ji}$ and $h \equiv h^j_j$ ($j$ summed). Moreover, the following four conditions will be imposed

$$\frac{\partial \psi^j_i}{\partial x^j} = 0, \tag{9}$$

where $x^j$ is a four vector: $x^j \equiv (x^0, x, y, z) = (ct, \mathbf{r})$.

Insertion of (6), (8) and (9) into (5) yields the approximate expression

$$\eta^{kk} \frac{\partial^2 \psi^j_0}{\partial x^k \partial x^k} = -16\pi c^{-4} G T^j_0, \tag{10}$$

where index $j$ runs from 0–3 and index $k$ is summed from 0–3. The components of the energy-momentum tensor $T^j_0$ in (10) are approximately given by

$$T^j_0 \cong c\, v^j \rho, \tag{11}$$

where $v^j \equiv dx^j/dt$. The result (10) in combination with the approximated values $T^j_0$ of (11) correspond to the linearized Einstein equations (compare, e.g., with [7] and [11, pp. 330–332]). It is noticed that apart from the four independent equations of (10) nine other equations follow from (5). The latter equations that contain second order velocity terms in $T^j_i$ (products of $v^i$ and $v^j$ ($i \neq 0$ and $j \neq 0$)) have been neglected. In the low velocity and weak field limit such an approximation is sufficiently accurate.

In order to obtain further analogy to the theory of electromagnetism, (10) can be rewritten by introduction of a scalar potential $\varphi$ defined as

$$\varphi \equiv \tfrac{1}{4} c^2 \psi^0_0 \tag{12}$$

and a vector potential $\mathbf{A}$ whose components are given by

$$A^\alpha \equiv \tfrac{1}{4} \beta\, c^2\, G^{-\frac{1}{2}} \psi^\alpha_0, \tag{13}$$

where $\alpha$ (= $x, y$ or $z$) is not summed.

Choosing $i = 0$ in (9) followed by insertion of (12) and (13) into (9) yields

$$\nabla \cdot \mathbf{A} + \beta\, c^{-1} G^{-\frac{1}{2}} \frac{\partial \varphi}{\partial t} = 0, \tag{14}$$

where $\nabla \equiv (\frac{\partial}{\partial x}, \frac{\partial}{\partial y}, \frac{\partial}{\partial z})$ is the Laplace operator. This relation is analogous to the Lorentz condition. Note that this result depends on the particular choice of the definitions (12) and (13). In addition, combination of (10)–(14) leads to two second order differential equations, one for the scalar potential $\varphi$ and one for the vector potential **A**, respectively

$$\nabla^2 \varphi - c^{-2} \frac{\partial^2 \varphi}{\partial t^2} = 4\pi G \rho, \tag{15}$$

$$\nabla^2 \mathbf{A} - c^{-2} \frac{\partial^2 \mathbf{A}}{\partial t^2} = 4\pi \beta\, c^{-1} G^{1/2} \rho\, \mathbf{v}, \tag{16}$$

where $\mathbf{v} \equiv (v^x, v^y, v^z)$. Currently, the exact value of $\beta$ is unknown. For symmetry reasons between (15) and (16) a value of $\beta = +1$ may be adopted. In order to demonstrate the influence of $\beta$ on other formulas, the general factor $\beta$ will be preserved.

In further analogy to electromagnetism, the gravitomagnetic field **B** may now be defined from **A**

$$\mathbf{B} \equiv \nabla \times \mathbf{A} \tag{17}$$

and a gravitational field **g** from $\varphi$ and **A** as

$$\mathbf{g} \equiv -\nabla \varphi - \beta\, c^{-1} G^{1/2} \frac{\partial \mathbf{A}}{\partial t}. \tag{18}$$

Note that the fields **B** and **g** are *not* mutually analogous. As will be discussed below, the definition of **B** in (17) implies that gravitomagnetic monopoles do not exist. Owing to the definition of **A** in (13), the gravitomagnetic field **B** from (17) possesses the dimension of a magnetic induction field, but other choices for **A** or **B** are also possible. For example, Weinberg [6] choose a dimensionless vector potential $\zeta$ defined in his equation (9.1.61). It can be shown that the relation between the vector potentials $\zeta$ and **A** is given by

$$\zeta = 4\beta^{-1} c^{-2} G^{1/2} \mathbf{A}. \tag{19}$$

From (14)–(18), by making use of standard vector-analysis, one can then derive the following set of four differential equations for sources in vacuum

$$\nabla \times \mathbf{B} = -4\pi \beta\, c^{-1} G^{1/2} \rho\, \mathbf{v} + \beta\, c^{-1} G^{-1/2} \frac{\partial \mathbf{g}}{\partial t}, \tag{20}$$

$$\nabla \cdot \mathbf{g} = -4\pi G \rho, \tag{21}$$

$$\nabla \times \mathbf{g} = -\beta^{-1} c^{-1} G^{1/2} \frac{\partial \mathbf{B}}{\partial t}, \tag{22}$$

$$\nabla \cdot \mathbf{B} = 0. \tag{23}$$

Especially for the choice $\beta = +1$, the analogy between the gravitomagnetic equations (20)–(23) and the four differential equations describing electromagnetic phenomena and known as the Maxwell equations is striking. In the static case (21) and (22) lead to Newton's well-known gravitation law. The striking analogy between this law and Coulomb's law has been noted since the beginning of physics. Therefore, the adopted analogy between the gravitational field **g** of (18) and the corresponding definition of the electric field **E** may also be valid. It is noticed that neither the Maxwell equations nor the gravitomagnetic equations are completely symmetrical. This asymmetry stems in the latter case from the definition of the gravitomagnetic field **B** = **B**(gm) in (17) implying relation (23). In the electromagnetic case the asymmetry is due to the corresponding definition of the electromagnetic magnetic induction field $\mathbf{B}(em) \equiv \nabla \times \mathbf{A}(em)$ implying the Maxwell equation $\nabla \cdot \mathbf{B}(em) = 0$.

The derivation of Maxwell-type gravitational equations has a long history, dating back to Heaviside already (see, e.g., [12]). A number of attempts to derive equations like (20)–(23) have been discussed, e.g., in ref. [9]. Deductions of the gravitomagnetic equations from the Einstein equations have been given by Peng [7], Biemond [8, 9], Mashhoon [13], Pascual-Sánchez [14], Ruggiero and Tartaglia [15] and others.

In the stationary case the gravitomagnetic field $\mathbf{B}$ can be obtained from the simplified equation (20) and (23)

$$\nabla \times \mathbf{B} = -4\pi \beta c^{-1} G^{\frac{1}{2}} \rho \mathbf{v} \text{ and } \nabla \cdot \mathbf{B} = 0. \qquad (24)$$

Since $\nabla \cdot \mathbf{B} = 0$, the field $\mathbf{B}$ can be derived from the vector potential $\mathbf{A}$ of (17). Equation (24) may imply that neutral matter moving either translationally or rotationally acts as a source of gravitomagnetism. In this work we will only deal with rotationally generated gravitomagnetism, implying $\mathbf{v} = \boldsymbol{\omega} \times \mathbf{r}$ ($\boldsymbol{\omega}$ is the angular velocity of a mass element $\rho\, dV$ and $\mathbf{r}$ is the distance from the rotation axis to this mass element). Translationally generated gravitomagnetism has shortly been discussed in [9].

For a massive rotating sphere with angular momentum $\mathbf{S}$ the following expression satisfies to (24)

$$\mathbf{A} = \tfrac{1}{2} \beta c^{-1} G^{\frac{1}{2}} \mathbf{S} \times \nabla 1/R, \qquad (25)$$

where again $\mathbf{S} = I \boldsymbol{\omega} = 2/5\, f\, m\, r_0^2\, \boldsymbol{\omega}$. In deriving (25) it has been assumed that the radius $r_0$ of the sphere is small in comparison with the distance $R$ from the centre of the sphere to the field point where the field $\mathbf{A}$ is observed. Utilizing (4), calculation of the gravitomagnetic field $\mathbf{B}$ from (17) yields

$$\mathbf{B} = -\tfrac{1}{2} \beta c^{-1} G^{\frac{1}{2}} \nabla(\mathbf{S} \cdot \nabla 1/R) = \nabla(\mathbf{M} \cdot \nabla 1/R). \qquad (26)$$

Further evaluation of (26) yields the gravitomagnetic field of (3). Equation (26) represents the field from the gravitomagnetic dipole moment $\mathbf{M} = \mathbf{M}(gm)$ of (4). Note that neither the sign nor the value of $\beta$ in (4) follow from the derivation. If the field $\mathbf{B}$ of (26) may be identified as a common magnetic induction field, then equation (4) may be denoted as the so-called Wilson-Blackett law [9].

The agreement between the observed magnetic induction field $\mathbf{B}(obs)$ of a rotating body and $\mathbf{B}(gm)$ from (4) (choosing $\beta = +1$) is reflected by the effective parameter $\beta_{eff}$

$$\mathbf{B}(obs) \equiv \beta_{eff}\, \mathbf{B}(gm). \qquad (27)$$

For $\beta_{eff} = +1$ full agreement occurs. For a series of about fourteen rotating bodies ranging from metallic cylinders in the laboratory to moons, planets, stars and the Galaxy an absolute value of $\beta_{eff} = 0.076$ has been found from a weighted least-squares fit to the data (see for details ref. [9] and references therein). Likewise, for fourteen slowly rotating, accreting, X-ray emitting, binary pulsars values of $\beta_{eff}$ lying in between 0.10 and 55 could be calculated [10]. Although the agreement with the gravitomagnetic prediction $\beta_{eff} = 1$ is only approximate, the found order of magnitude of $\beta_{eff}$ for so *many, strongly different*, rotating bodies may reflect the basic validity of the Wilson-Blackett law.

As pointed out earlier [9], moving electric charge in the magnetic field from gravitomagnetic origin may cause an additional magnetic field from electromagnetic origin. Moreover, it is noticed that the predicted magnetic field generated by a rotating mass also bearing a charge is generally much smaller than the magnetic field generated by the charge. Therefore, it is usually extremely difficult to isolate the gravitomagnetic contribution from an observed magnetic field. For these reasons, magnetic fields from gravitomagnetic origin may have escaped detection so far.

Some remarks with respect to equation (23) can be made. Equation (23) follows from (17), whether the gravitomagnetic field **B** is identified as a magnetic induction field or not. As a consequence of (23), magnetic monopoles from gravitomagnetic origin do not exist. Analogously, the Maxwell equation $\nabla \cdot \mathbf{B}(em) = 0$ excludes the existence of magnetic monopoles from electromagnetic origin. If **B** in (23) is equivalent to a magnetic induction field due to a charge, then (23) and $\nabla \cdot \mathbf{B}(em) = 0$ may explain why no magnetic monopoles of any kind have been detected so far. For a discussion of the present status of magnetic monopole research, see, e.g., Groom [16].

## 3. Gravitomagnetic precession of a gyroscope

Starting from the principle of general covariance the following equations of spin motion can be derived [6, chapters 5, 9]

$$\frac{dS_i}{d\tau} = \Gamma^j{}_{ik} S_j \frac{dx^k}{d\tau}, \tag{28}$$

where the quantities $\tau$, $S_i$ and $\Gamma^j{}_{ik}$ have their usual meaning. The index $i$ runs from 0–3, whereas the indices $j$ and $k$ are summed from 0–3. If we set $i = \alpha$, multiply both sides of (28) with $d\tau/dt$ and eliminate $S_0$, the gravitomagnetic contribution can be separated off from this equation. By considering only the terms with $k = 0$ in (28), one obtains

$$\frac{dS_\alpha}{dt} \cong c\, \Gamma^\beta{}_{\alpha 0} S_\beta, \tag{29}$$

where index $\alpha$ runs from 1–3 and index $\beta$ is summed from 1–3. It is noticed that the terms of the so-called geodetic precession have not been included in (29). We shall not deal with them in this work.

Utilizing (8), introduction of $A^\alpha$ from (13) into $\Gamma^\beta{}_{\alpha 0}$, followed by insertion of the result into (29), yields in the weak field limit

$$\frac{dS_\alpha}{dt} = 2\, \beta^{-1} c^{-1} G^{\frac{1}{2}} \left( \frac{\partial A^\beta}{\partial x^\alpha} - \frac{\partial A^\alpha}{\partial x^\beta} \right) S_\beta - \tfrac{1}{2} \frac{\partial g_{\beta\alpha}}{\partial t} S_\beta. \tag{30}$$

Evaluation of $g_{\beta\alpha}$ in (30) by introduction of the scalar potential $\varphi = -Gm/R$ then leads to the following three dimensional vector form for $d\mathbf{S}/dt$ ($\mathbf{S} \equiv S_x, S_y, S_z$)

$$\frac{d\mathbf{S}}{dt} = -2\, \beta^{-1} c^{-1} G^{\frac{1}{2}} (\nabla \times \mathbf{A}) \times \mathbf{S} - c^{-2} \frac{\partial \varphi}{\partial t} \mathbf{S}. \tag{31}$$

Note that the terms on the right hand side of (31) correspond to those of Weinberg [6] (compare with his equation (9.6.5)). Substitution of **B** from (17) into (31) finally leads to the following gravitomagnetic contribution to $d\mathbf{S}/dt$ (the term in $\partial\varphi/\partial t$ in (31) not depending on **A** will not further be considered here)

$$\frac{d\mathbf{S}}{dt} = -2\, \beta^{-1} c^{-1} G^{\frac{1}{2}}\, \mathbf{B} \times \mathbf{S} = \mathbf{\Omega} \times \mathbf{S}. \tag{32}$$

From this relation the gravitomagnetic precession rate **Ω** of (2) follows.

Related derivations of (1), (2) and (32) have previously been given by Weinberg [6] and Peng [7]. An alternative way to deduce them is to start from the gravitomagnetic force density **f** [9]

$$\mathbf{f} = 4\,\beta^{-1}\,c^{-1}\,G^{\frac{1}{2}}\,\rho\,\mathbf{v} \times \mathbf{B}. \tag{33}$$

The latter equation can be deduced by combining (7) and (13). Utilizing (33) and following the same line of reasoning given by Landau and Lifshitz [11] for the electromagnetic analogue, the formulas (1), (2) and (32) can be calculated [9].

If the gravitomagnetic field $\mathbf{B} = \mathbf{B}(gm)$ from (3) generated by rotating, electrically neutral mass will be incompatible with a magnetic induction field $\mathbf{B} = \mathbf{B}(em)$ from electromagnetic origin, then substitution of $\mathbf{B} = \mathbf{B}(gm)$ from (3) and (4) into (2) leads to the standard result for the precession rate of $\mathbf{\Omega}$ of (1). For a gyroscope in the Gravity Probe B moving in a 649 km orbit above the Earth ($S = 5.87 \times 10^{40}$ g.cm$^2$.s$^{-1}$, $R = 7.027 \times 10^8$ cm) the integrated value of $\Omega$ would then be equal to the standard value of 40.9 milliarc-seconds per year (mas/yr) [1–3]. It is noticed that within experimental accuracy the slight eccentricity $e$ ($e = 0.0014$, see [1]) of the orbit of GP-B does not affect this result.

The (absolute) value of the gravitomagnetic dipole moment $M(gm)$ of the Earth can be calculated from (4). Choosing $\beta = +1$ and substitution of $S = 5.87 \times 10^{40}$ g.cm$^2$.s$^{-1}$ into (4) yields $M(gm) = 2.53 \times 10^{26}$ G.cm$^3$. The observed magnetic moment $M(obs)$ of the Earth, however, is equal to $7.91 \times 10^{25}$ G.cm$^3$. Combination of (3), (4) and (27) then yields a value of 0.31 for $\beta_{eff}$. So, in this case circulating charge in the Earth may *weaken* the magnetic field caused by gravitomagnetism.

If the field $\mathbf{B}(gm)$ shows the same behaviour as a field $\mathbf{B}(em)$ generated by moving charge, then the precession rate $\mathbf{\Omega}$ of (2) may be determined by the total magnetic induction field $\mathbf{B}(tot) = \mathbf{B}(gm) + \mathbf{B}(em)$. The field $B(gm)$ can be calculated from a combination of (3) and (4). From (3) the polar field $B_p(gm)$ at a gyroscope in the Gravity Probe B ($R = 7.027 \times 10^8$ cm) can be calculated to be 1.46 G, whereas the orbital-averaged magnetic field is equal to ¼ $B_p(gm) = 0.36$ G. However, the gyroscopes in the GP-B experiment have been shielded against external dc magnetic fields of order $10^{-7}$ G [1–3], so that the residual integrated value for $\Omega$ from (2) could be equal to $1.1 \times 10^{-5}$ mas/yr. Although the accuracy of the GP-B experiment is expected to be better than 1% [1, 3], or 0.4 mas/yr, this small value is outside observational reach. Thus, the observed precession rate $\Omega$ may clarify the nature of the gravitomagnetic field $B(gm)$.

A final remark may be justified here on the exact value of $\beta$. This work suggest a value of $\beta = +1$ based on the discussion of (16) and the gravitomagnetic equations (20)–(23). In literature various choices for the dimensionless constant $\beta$ have been made implicitly. For example, Weinberg [6] applies $\beta = +4$ (compare his equations (9.1.66) and (9.6.5) with equations (14) and (31), respectively), Mashhoon [13] and Ruggiero and Tartaglia [15] use $\beta = +2$, Peng [7] and Pascual-Sánchez [14] use $\beta = +1$. Note that a negative value for $\beta$ is also mathematically possible. For example, Heaviside applies a value of $\beta = -1$ [12], which makes the direction of $\mathbf{\Omega}$ with **B** in (2) coincide.

## 4. Conclusions

If the gravitomagnetic field $\mathbf{B} = \mathbf{B}(gm)$ from (3) generated by rotating, electrically neutral mass will be incompatible with a magnetic induction field $\mathbf{B} = \mathbf{B}(em)$ from electromagnetic origin, then substitution of $\mathbf{B} = \mathbf{B}(gm)$ from (3) and (4) into (2) leads to the standard result for the precession rate $\mathbf{\Omega}$ of (1). For a gyroscope in the Gravity Probe B moving in a 649 km orbit above the Earth ($S = 5.87 \times 10^{40}$ g.cm$^2$.s$^{-1}$, $R = 7.027 \times 10^8$ cm) the integrated value of $\Omega$ would then be equal to 40.9 mas/yr [1–3].

The alternative gravitomagnetic precession rate $\mathbf{\Omega}$ of (2) may also be deduced from general relativity. If the gravitomagnetic field $\mathbf{B} = \mathbf{B}(gm)$ of (3) from gravitational origin behaves like a magnetic induction field from electromagnetic origin, **B**(em), then the precession rate **Ω** of (2) will be determined by the total magnetic induction field **B**(tot) = **B**(gm) + **B**(em). However, the four gyroscopes in the Gravity Probe B have been shielded against external dc magnetic fields of order $10^{-7}$ G, so that the residual integrated value of the precession rate $\Omega$ from (2) could be equal to $1.1 \times 10^{-5}$ mas/yr, outside the observational reach of GP-B.

The approximate validity of the so-called Wilson-Blackett law (4) for rotating massive bodies [8–10] is an indication in favour for the identification of $B$(gm) of (3) as a common magnetic induction field. Discrepancies between the values of $B$(gm) following from (4) and the observed field $B$(obs) may be attributed to interfering effects from electromagnetic origin [9, 10].

Nayeri and Reynolds [17] also discussed the gravitomagnetic precession in the context of the five-dimensional model of Randall and Sundrum. This model can be considered as an extension of general relativity. From the two-brane scenario of their model a value of $4 \times 10^{-31}$ mas/yr for the gravitomagnetic precession rate $\Omega$ can be calculated, far outside the observational reach of GP-B. Note that all discussed predictions for $\Omega$ (the standard value $\Omega$ = 40.9 mas/yr, the value $\Omega \cong 0$ from Nayeri and Reynolds [16] and our value $\Omega \cong 0$) start from general relativity as a basis.

It is to be expected that the results of the Gravity Probe B will discriminate between the standard gravitomagnetic precession rate $\Omega$ from (1) and the alternative predictions $\Omega \cong 0$ discussed in this work. Thus, the true nature of the gravitomagnetic field **B** in (3) may become more manifest. If the latter field appears to be equivalent to the magnetic field of electromagnetic origin, then mass and charge are intimately connected by magnetism. Such a result would be an important step towards further unification of the theories of general relativity and electromagnetism (compare with, e.g., [18]).

## Acknowledgement

I should like to thank my son Pieter for technical realisation and publication of this paper.